\documentclass[10pt]{article}

\usepackage{amssymb}
\usepackage{natbib}

\usepackage[figuresright]{rotating}

\usepackage[bookmarks=false]{hyperref}
    \hypersetup{colorlinks,
      linkcolor=blue,
      citecolor=blue,
      urlcolor=blue}

\usepackage{multicol}
\usepackage{mdframed}

\usepackage[margin=2cm]{geometry}

\title{An agent-based model for modal shift in public transport}

\author{Thibaut Barbet$^{1}$, Amine Nacer-Weill$^{1}$, Changtao Yang$^{1}$ and Juste Raimbault$^{2,\ast}$\medskip\\
$^{1}$ Ecole des Ponts ParisTech\\
$^{2}$ Center for Advanced Spatial Analysis, University College London\medskip\\
$^{\ast}$ \texttt{juste.raimbault@polytechnique.edu}
}
\date{}

\begin{document}

\maketitle

\begin{abstract}
Modal shift in public transport as a consequence of a disruption on a line has in some cases unforeseen consequences such as an increase in congestion in the rest of the network. How information is provided to users and their behavior plays a central role in such configurations. We introduce here a simple and stylised agent-based model aimed at understanding the impact of behavioural parameters on modal shift. The model is applied on a case study based on a stated preference survey for a segment of Paris suburban train network. We systematically explore the parameter space and show non-trivial patterns of congestion for some values of discrete choice parameters linked to perceived wait time and congestion. We also apply a genetic optimisation algorithm to the model to search for optimal compromises between congestion in different modes.\\\medskip

\end{abstract}

\section{Introduction}

Disruptions in public transport networks are not rare events, often worsened by an increased complexity of these networks and their management \citep{dekker2018next}. Network resilience is then tightly linked to patterns of modal shift \citep{stamos2015impact}, and a better understanding of these is crucial both from a theoretical and operational viewpoint. In that context, users have to make decisions in a limited time and under partial information \citep{lyons2006role}. The study of modal shift under disruption is therefore improved when taking into account users behavior and a detailed representation of users cognition \citep{brisbois2010processus}. More particularly, the role of information provided in real time can in some cases become crucial, as rerouting may in fact increase the congestion in other parts of the network and ultimately increase the average travel time \citep{chatterjee2002driver,chorus2006travel}.

% check https://www.mdpi.com/2071-1050/11/6/1765

To study mechanisms linking information given to users with modal shift, and more generally the evolution of multi-modal network flows under disruptions, agent-based modeling has been highlighted as a relevant approach \citep{leng2020role}. For example, \cite{leng2020issue} show that the issue time of information has a significant impact on the total congestion. Agent-based models are used in similar studies of modal share, such as by \cite{baindur2011agent} for freight. \cite{ambra2019should} include the reaction to disruptions in a multimodal agent-based model. \cite{raney2003agent} describe an application of the MATSim model, which is a data-driven agent-based and activity-based transport model, to a large sample of Switzerland transport network. The MATSim model can be applied at large scales in a reproducible manner, such as in the case of Ile-de-France illustrated by \cite{horl2020reproducible}.

Large transport agent-based models such as MATSim however require an extensive parametrisation on real data, are difficult to systematically validate given their runtime and large parameter space, and despite their high modularity can be tuned only to some extent regarding a precise description of users behavior in public transport and their interactions with a network disruption. This paper therefore proposes to introduce a simple and stylised agent-based model to understand the role of information and user behavior in modal shift under disruptions in public transport. With a reduced computational complexity but also parameter space to explore, under limited requirements for data, such a model can be used to systematically explore if some qualitative stylised facts are robustly found under different scenario, and possibly extended into a decision-making tool. Our contribution is threefold: (i) we contextualise the construction of the model with a reduced stated preferences survey, applied on a case study in the Parisian public transport network on a specific segment often subject to disruption; (ii) we introduce the simple agent-based model as an open-source tool, which can be extended or modified to test concurrent hypotheses on user behavior; (iii) we proceed to a systematic exploration of the model parameter space, unveiling some non-linear patterns and a counter-productive reaction of users to the disruption in terms of congestion for some scenarios.

The rest of this paper is organised as follows: we first describe an exploratory stated preference survey; we then describe the agent-based model and results of numerical experiments. We finally discuss potential applications and developments.

\section{Stated preferences survey}

We first proceeded to a small size stated preferences survey \citep{kroes1988stated}, in order to have a qualitative overview of processes needed in the model and to have a case study for model application. \cite{martin2016strategies} have indeed shown that there exists a high heterogeneity in user reaction to disruptions. We choose to study the line A of the \emph{RER} suburban train in Paris, which has the highest load in the region and has a non-negligible frequency of disruption. We focus on a subpart of the network, in order to sample users which have a higher chance of realising a given origin-destination pattern, and for which several modal alternatives exist. Therefore, the segment \emph{Etoile-La D{\'e}fense} was chosen, as it features alternative trips with the \emph{Metro Line 1} and several bus lines.

%%%%%%%%%%%%
\begin{figure}%[t]\vspace*{4pt}
\centerline{\includegraphics[width=\linewidth]{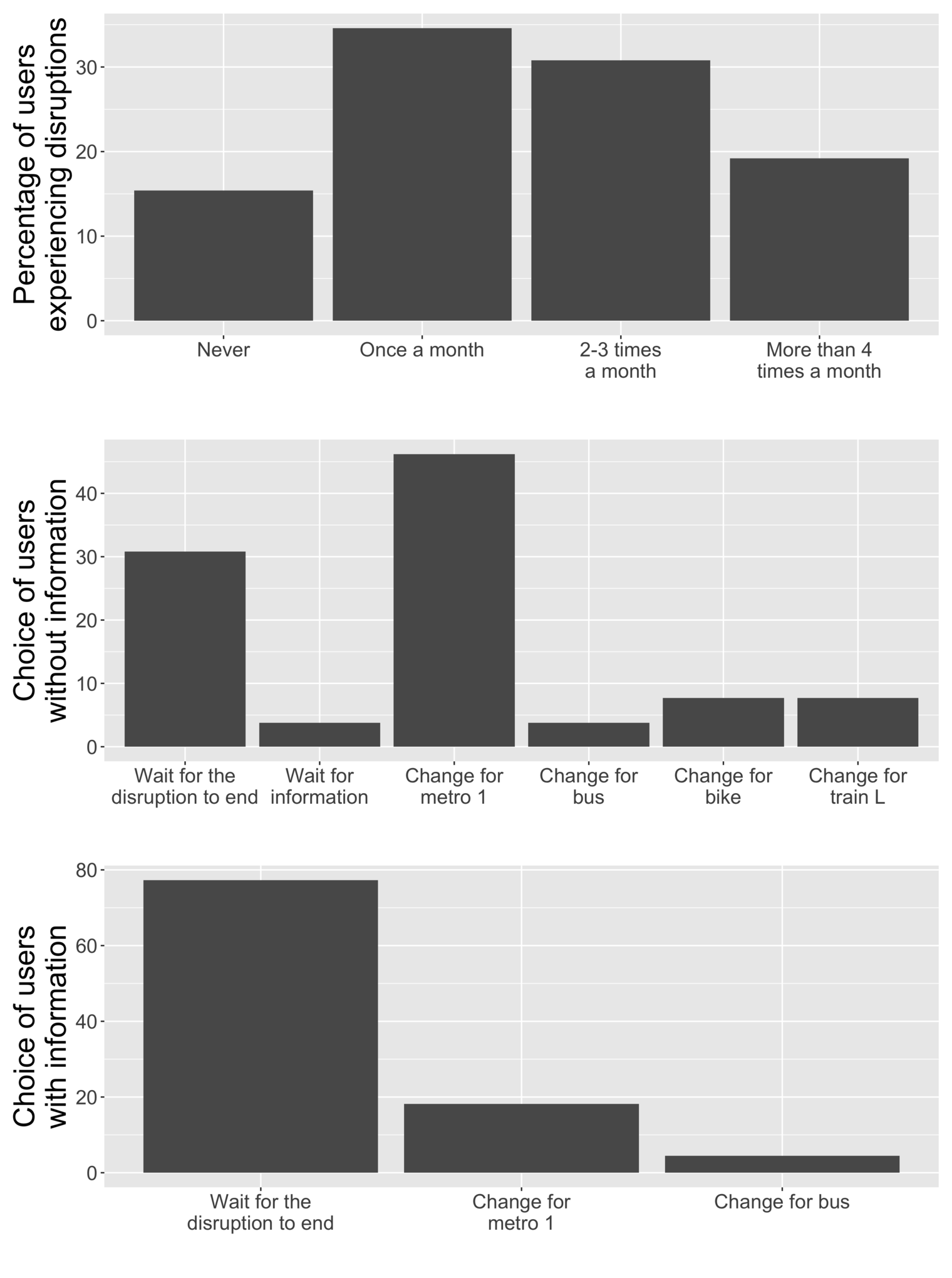}}
\caption{Results of the stated preference survey. \textit{(Top)} Number of times a month users experience a disruption; \textit{(Middle)} Behavior of users when no information is given; \textit{(Bottom)} Behavior of users when a 10 minutes traffic stop is announced.\label{fig:fig1}}
\end{figure}
%%%%%%%%%%%%

We surveyed a total of $N=48$ users, among which a sub-sample of $N=27$ were regular users for which the full questionnaire was given. Surveys were realised on peak hours (between 7am and 9am) of weekdays in January 2021, on the platform of the \emph{Etoile} station. The exact survey which was used (in French) is available on the open repository of the project (see model implementation below).

Main results which can inform our model context are shown in Fig.~\ref{fig:fig1}. We first confirm that disruptions are indeed frequent on the line, as around 50\% of users experience a disruption at least twice a month. We then find that under no information, user behavior is heterogeneous and a large part shifts to alternative itineraries. Finally, when users are informed of a 10 minutes disruption, on the contrary more than 75\% decide to wait on place. Therefore, the model must include various modal shifts, but also take into account how perceived wait time will influence user choice.

This survey was designed in an exploratory manner as a pilot study, and could not be continued on a larger scale for various practical reasons. Our sample size thus remains relatively small and this part of the study must be considered as qualitative. Significantly larger sample sizes are needed to obtain robust inferential results, for example to forecast travel demand \citep{horvath2015estimation} or to estimate discrete choice parameters \citep{bliemer2009efficiency}. In our case, we consider the simulation model as a tool to overcome such data limitations and still study stylised behavioural processes using qualitative insights from our survey data.

\section{Model description}

The agent-based model is based on two types of agents: users and RER trains. The simulated segment of the network is included for the main mode (RER train) and alternative modes (metro, bus, taxi, bike, walking). Main mode is simulated in full granularity, i.e. including train boarding, train capacity and train alighting. Other modes are simulated as queues with a given capacity. One simulation covers peak hours, which we take as a duration of 4 hours, with one-minutes time steps. The model is therefore stopped at $t_f = 240$. Arrival of users for each mode are simulated using stationary Poisson random draws.

More precisely, the model simulates sequentially the following processes at each time step.

\begin{enumerate}
	\item New users enters the network: for each mode, a fixed arrival rate $\lambda_i$ is used for a random Poisson law draw giving the number of new users entering the segment on this mode. Users entering the main RER mode are set on the platform as waiting, while users on other modes are queued at the start of the segment.
	\item Users waiting for a RER train on the platform evaluate a discrete choice utility difference (between waiting and shifting to an other mode) given by 
	\begin{equation} 
	\Delta U = \beta_c \cdot c + \beta_{\tau} \tau
	\end{equation}
	where $c$ is perceived congestion given by the current number of user waiting normalised by platform capacity (which corresponds in our case only to a parameter rescaling as we consider a single station); $\tau$ is perceived time that we assume proportional to user current travel time (time since departure of the user trip) and current waiting time between trains. $\beta_c$ and $\beta_{\tau}$ are thus behavioural parameters capturing the influence of perceived time and congestion on user behaviour. Note that a model generalisation with a broader scope (more stations for example) will have to introduce more accurate proxies for this processes or other behavioural aspects.
	\item These users can then switch mode with the corresponding discrete choice probability probability given by 
	\begin{equation}
	p = 1 / (1 + \exp \Delta U)
	\end{equation}
	and with fixed nested probabilities for the choice of the other mode once a shift has been decided;
	\item Next train possibly enter station (once the time interval $I$ between trains has elapsed) and users board at a given speed and given train capacity. Train capacity $C$ captures congestion upstream this segment, and is given as a maximal number of users which can board a train. In the meantime, users alight at the end of the segment if a train has arrived at the terminus station.
	\item Train queue on the segment is simulated, i.e. each train is advanced at its maximal speed, if the next slot is not occupied by a preceding train.
	\item Other modes are simulated as queues with a given user capacity: users enter the queue if it contains less users than its capacity; users in the queue are advanced according to the mode speed; arrived users are removed from the queue.
\end{enumerate}

Model results are assessed with indicators giving the average travel time and average congestion over the peak hour and all users. Congestion indicators can be computed for each mode separately.

The model has several parameters which can be set with realistic values (see setup below), while others remain variables of interest which influence on model behavior will be studied in numerical experiments. In particular, varying parameters are the behavioural parameters $\beta_c$ and $\beta_{\tau}$, time interval between train $I$ and train capacity $C$.

The discrete choice function does not include control variables such as differences between modes in ticket price, level of service, quality of information. These are implicitly taken into account in our model through the empirical values of modal shift observed in the survey, and could be included in a refined version of the model.

\section{Results}

\subsection{Implementation and model setup}

\begin{figure}%[t]\vspace*{4pt}
\begin{mdframed}
\centerline{\includegraphics[width=\linewidth]{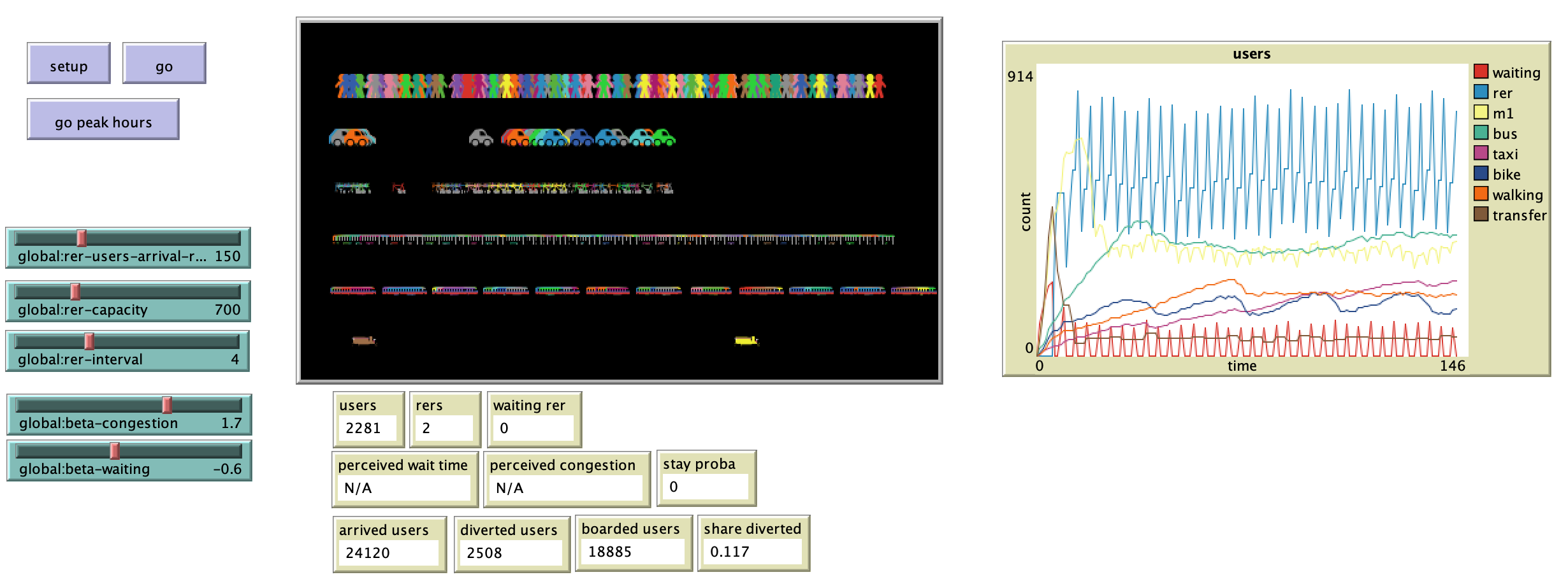}}
\end{mdframed}
\caption{Model interface. Main varying parameters are included as sliders, while mode queues are visualised in a spatialised manner and as a plot showing the load for each mode in time. Various numerical indicators are provided.\label{fig:fig2}}
\end{figure}

The model is implemented in NetLogo which is a platform and programming language specifically suited for such agent-based simulations \citep{tisue2004netlogo}. We show a visualisation of model interface in Fig.~\ref{fig:fig2}. The open-source code of the model and results is available as a git repository at \url{https://github.com/JusteRaimbault/ReportMasse}. We systematically explored model parameter space by using the OpenMOLE model exploration software \citep{reuillon2013openmole}, which allows embedding any type of model, provides a seamless access to high performance computing infrastructures, and integrates state-of-the-art model validation and exploration methods.

Several model parameters are set following real world values. Respective speed for each mode for example correspond to travel times for the segment in an uncongested setting. Mode capacities also have realistic values (for example for Metro 1, 700 users per train with 5 trains on the segment). We take a boarding speed for the RER of 1000 users per minute (MI09 trains have a total capacity of 2600 users and stop for not more than 1min30sec during peak hours). A train takes 4min for the segment. We assume a maximal waiting time in station of 2 minutes. We fix the user arrival rate $\lambda_{RER} = 100$ in experiments, to study borderline scenarios in terms of congestion. Finally, user transfer time between modes is taken as 5 minutes.

%%%%%%%%%%%%%
%%%%%%%%%%%%

\subsection{Parameter space exploration}

\begin{figure}%[t]\vspace*{4pt}
\centerline{\includegraphics[width=\linewidth]{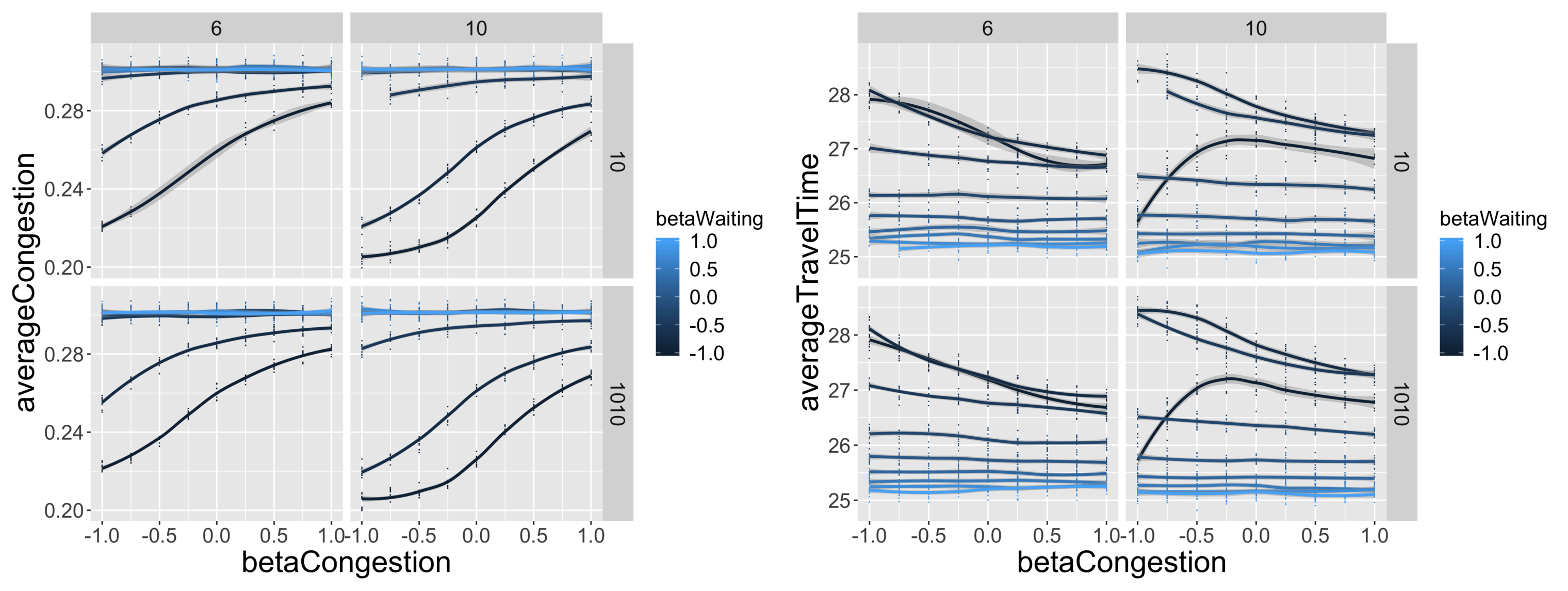}}
\caption{Model exploration results. \textit{(Left)} Average congestion as a function of $\beta_c$, for varying $\beta_{\tau}$ (colour), train interval (columns) and train capacity (rows); \textit{(Right)} Average travel time for same parameter values.\label{fig:fig3}}
\end{figure}

We run 10 stochastic replications of the model, for parameters $\beta_c,\beta_{\tau}$, train capacity $C$, and train arrival interval $I$ varying within a grid. This corresponds to $\simeq 24,000$ runs of the model. We show in Fig.~\ref{fig:fig3} the variation of average travel time and congestion as a function of $\beta_c$. We find an interesting non-linear behavior of travel time (Fig.~\ref{fig:fig3}, Right panel) for low values of $\beta_{\tau}$ (dark blue, right column), where travel time is maximal around a neutral position of users to congestion: in that context of a low tolerance to waiting, either a low or high tolerance to congestion are better than being indifferent. Larger values of $\beta_{\tau}$ induce a decreasing travel time as a function of $\beta_c$ (the more users are not tolerant to congestion, the lower the travel time), and similarly travel time is decreasing as a function of $\beta_{\tau}$. Regarding the congestion indicator (Left panel of Fig.~\ref{fig:fig3}), it increases with both discrete choice parameters but differently in the various disruption scenarios.

%%%%%%%%%%%%
%%%%%%%%%%%%

\subsection{Model optimisation}

We then propose to apply the model in an optimisation context. We use a genetic algorithm for multi-objective optimisation (more precisely the NSGA2 algorithm implemented into the OpenMOLE platform) to search for compromises between congestion in the train and congestion in alternative modes. We run the algorithm in a congested context (arrival rate $\lambda_{RER} = 100$, train capacity $C=500$ and train interval $I=5$), with free parameters $\beta_c$ and $\beta_{\tau}$. Optimisation objectives are the average congestion in the RER and the average congestion in other modes.

\begin{figure}%[t]\vspace*{4pt}
\centerline{\includegraphics[width=0.75\linewidth]{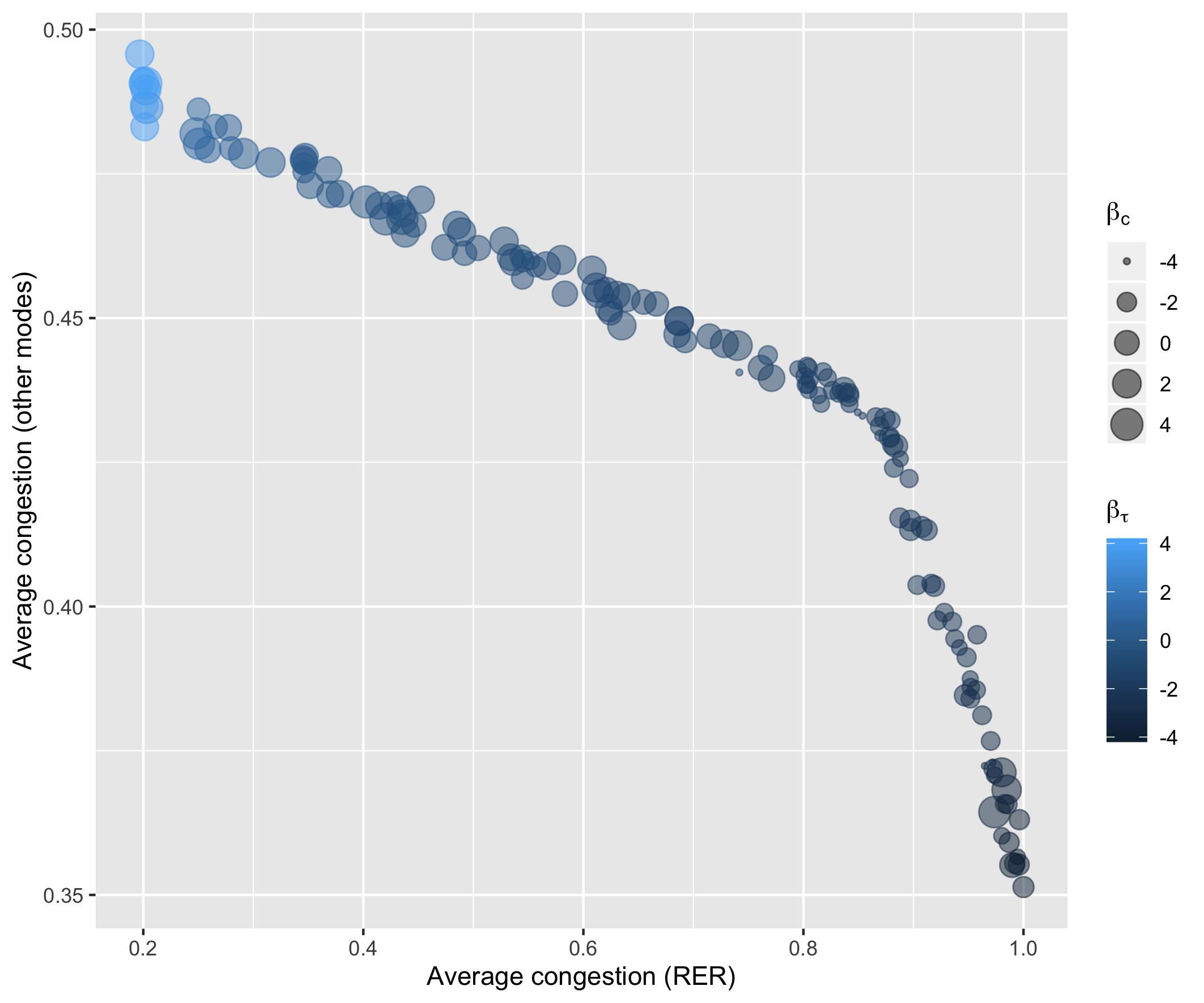}}
\caption{Model optimisation. We plot the Pareto front of compromise points between the two optimisation objectives. Point colour level gives $\beta_{\tau}$ while point size gives $\beta_c$.\label{fig:fig4}}
\end{figure}

Simulation results obtained with an algorithm population of $\mu = 200$ after 2000 generations are shown in Fig.~\ref{fig:fig4}. We find that acting on different dimensions of user behavior can mitigate the different dimensions of congestion. More precisely, a Pareto front is obtained between congestion in the RER and congestion in the other modes. Congestion in the RER can be minimised, at the detriment of other modes. This occurs when congestion and perceived time parameters have both high values. On the contrary, a low congestion in other modes implies a high tolerance of users and a high congestion in the RER. The shape of the Pareto front is interesting, witnessing two linear regimes. A good compromise is thus found at the breakpoint. At this point, users are rather tolerant to perceived time (negative $\beta_{\tau}$ values) but can have various behaviours regarding congestion. This shows that mitigating congestion can be achieved by acting on user behaviour, and that a compromise for the multimodal network can be found.

%%%%%%%%%%%%
%%%%%%%%%%%%

\section{Discussion}

Our simulation results show how this model can be applied to explore complex congestion patterns in the stylised network following interactions between different components of user behavior. Our model can be applied to the simulation of disruption events, and possible scenarios to improve network resilience. It can also be applied to test interventions to mitigate congestion in targeted modes. Its advantage in comparison to more complicated models relies on its simplicity, which allows systematic explorations of the parameter space and more advanced numerical experiments as we illustrated above.

Several developments can furthermore be considered. The behavioural model is rather simple, and as the implementation is generic, other functions or more complicated processes can be tested. The variables we used in the discrete choice utility are only proxies of perceived information, and how to account for the information effectively received by the users depending on the context remains an open question \citep{gao2018trip}. The disutility due to perceived travel time is in our case simply the sum of previous travel time and announced waiting time: studying other functions and aspects is an other interesting development of the model. For example, including congestion within the trains may be crucial regarding recent health concerns linked to crowding in closed spaces \citep{raimbault2021estimating}.

Besides, the possible mitigation of congestion through interventions acting on the provided information may raise ethical issues - how to ensure an overall optimisation while remaining fair to all users: it was shown in other modes such as car travel that inequalities in congestion may rise from path shift due to live information \citep{cabannes2017impact}. We also assumed on this issue of access to information that users did not necessarily had access to some synthetic information on all alternatives (with a web application for example) and that there was no unified mobility service. Exploring the impact of such aspects on behavioural and congestion patterns is also a potential refinement of the model.

Finally, a more data-driven version of the model, including estimated discrete choice parameters, would be an interesting path to explore, assuming that a further survey and data collection is possible. This may also imply to include control variables in discrete choices utilities to account for differences between modes for example in terms of price or level of service.

\section{Conclusion}

The study of modal shift following a disruption in public transport networks is an important aspect to ensure resilience and a high level of service. We introduced in this paper the basis for a simple agent-based model focused on user decision based on perceived congestion and waiting time. We showed how the model can be used to explore the interaction between different behavioural components of users, and to optimise the congestion in different modes. Our model is open source and aimed at being extended and applied to other case studies. Systematic model exploration and validation should provide both practical results in terms of transport management, but also more general results on generic and specific processes in modal shift.

%% References

%The citation must be used in following style: \cite{article-minimal} \cite{article-full} \cite{article-crossref} \cite{whole-journal}.

%

%\bibliographystyle{elsarticle-harv}
%\bibliography{biblio.bib}

\clearpage

\end{document}